
\input vanilla.sty
\font\tenbf=cmbx10

\font\ninebf=cmbx9
\font\ninerm=cmr9
\font\nineit=cmti9

\font\eightrm=cmr8
\font\eightit=cmti8

\TagsOnRight
\hsize=15 truecm
\vsize=22 truecm
\hoffset =1.0 truecm
\parindent=15pt
\baselineskip=12pt
\def\qed{\hbox{${\vcenter{\vbox{
    \hrule height 0.4pt\hbox{\vrule width 0.4pt height 6pt
    \kern5pt\vrule width 0.4pt}\hrule height 0.4pt}}}$}}

\line{\eightrm  LANDAU-TMP-4-93\hfil}
\line{\eightrm  hep-th/9307127\hfil}
\vglue 5pc
\baselineskip=13pt
\centerline{\tenbf  LATTICE $W$ ALGEBRAS AND}
\centerline{\tenbf  QUANTUM GROUPS.
\footnote"$^{+}$"
{\eightrm\baselineskip=10pt Talk presented on the 3 International
Conference on Mathematical Physics, String Theory and Quantum
Graviry, Alushta 1993.
}
}
\vglue 24pt
\centerline{\eightrm Ya.P. Pugay
}
\baselineskip=12pt
\centerline{\eightit Landau Institute for Theoretical Physics}
\baselineskip=10pt
\centerline{\eightit 142432 Chernogolovka, Russia}
\vglue 20pt
\centerline{\eightrm ABSTRACT}
{\rightskip=1.5pc
\leftskip=1.5pc
\eightrm\baselineskip=10pt\parindent=1pc
We represent Feigin's construction [22] of lattice W algebras
and give some simple results: lattice Virasoro and $W_3$ algebras.
For simplest case $g=sl(2)$ we introduce whole
$U_q(sl(2))$ quantum group on this lattice.
We find simplest two-dimensional
module as well as exchange relations and define lattice Virasoro
algebra as algebra of invariants of $U_q(sl(2))$.
Another generalization is connected with lattice integrals of motion
as the invariants of quantum affine group $U_q(\hat{n}_{+})$. We show
that Volkov's scheme leads to the system of difference equations
for the function from non-commutative variables.
\vglue 13pt
\rm\baselineskip=15pt\parindent=1pc
\line{\tenbf 0. Introduction. \hfil}
\vglue 5pt
In this talk I would like to give a brief introduction
to the Feigin [22] construction of lattice
$W$ algebras and represent some simple results. More
complete consideration can be find in the forthcoming work
[36].

In 1985 Alexander Zamolodchikov [1] investigated the possibility
of existence of new additional infinite symmetries in
the context of two-dimensional Conformal Field Theory [4],
or, equivalently, the existence of a primary field
with conformal dimension $(s,0)$ or $(0,s)$. (Hereafter we
consider only holomorphic part.)
By direct use of bootstrap principle he proved that there might
exist primary
field $W_3$ with conformal dimensions $(3,0)$. Due to the equation
$$
\partial_{\bar{z}}W_3(z)=0 \quad ,
$$
it is the conserved current which generates additional infinite
symmetry, while algebra $(W_3,T)$ (where $T$ is the stress-energy
tensor), is a quadratic one. Namely, operator product expansion of two $W_3$
currents includes quadratic term on $T$.
In the past few years
considerable progress has been made in an understanding
of the deep structures underlying these algebras
(see for example refs.[2,5-15,20,21]) as well as its classical
limits [16-19].
It was shown
in the works [2,6,11,12,14]
that $W$ algebras can be considered as the result of quantum
Drinfel'd-Sokolov reduction and the fact that generators
of $W$ algebras commute with screenings operators can be
taken as the definition of $W$ algebras. Namely, such as
screening operators constitute the nilpotent part of the
quantum group [24],
mathematically W algebra is the algebra of invariants of this group.
Lukyanov and Fateev [5-8] found that such invariants are given by
quantum generalization of Miura transformation. We give the
lattice version of this picture. We hope that lattice
construction can clear up the intrinsic fiatures of $W$
algebras. Our consideration is somewhat different from
a number of another works appeared in connection with
lattice current algebras [23-31].

Plan of this talk is following:

1. Feigin's construction of lattice $W$ algebras.

2. Examples: Virasoro and $W_3$ on the lattice.

3. Generalization: $U_q(sl(2))$.

4. 2D-modules of $U_q(sl(2))$. Exchange algebra. Virasoro algebra.

5. Generalizations: affine case and integrals of motion.

\voffset=2\baselineskip
\vglue 12pt
\line{\tenbf 1. Feigin's construction of lattice $W$ algebras. \hfil}
\vglue 5pt
\line{\tenbf 1.1 W algebra in continious theory.\hfil}
Consider bosonic representation of conformal field theory
in terms of free scalar fields $\phi^{b}$ ($b=1,...,r$ with expansion
$$
T(\phi^{a}(\zeta)\phi^{b}(z))= {{\delta^{ab}}\over k} \log(\zeta-z)
+O(\zeta-z) ,
\tag 1.1.1$$
Let $g$ be a simple Lie algebra with simple roots $\alpha_i$
(i=1,...,r=rank(g)) and Cartan decomposition $g=n_{-}\oplus h \oplus n_{+}$.
Following to Feigin and Frenkel [12] one can give the following definition:
\vglue 5pt
\noindent
{\tenbf Definition.}Vacuum representation of $W$ algebra associated with
simple Lie algebra $g$ has
a realization as the intersection of the kernels of screening operators
$$
S_{\alpha_{i}}=\oint{:[exp(-\alpha_{i}\phi)]:{dz}} \quad , \quad 1<i<r.
\tag 1.1.2$$

We have by this means in the continious theory
$$
Wg \sim Ker(S_{\alpha_1})\cap Ker(S_{\alpha_2}) \cap Ker(S_{\alpha_3}) \quad
... \quad .
$$
Lukyanov and Fateev found that bosonic realization of $W$ is given by quantum
Miura transformation which in the case of $Wsl(n)$ has the following
form [5-8]:
$$
\partial^{n}+\sum^{n}_{l=1} W_{l}(x) \partial^{n-l}=
:[\prod^{n}_{i=1}(\partial-\vec{h}_i\vec{\phi}'(x))]: \quad .
\tag 1.1.3$$
and
$\vec{h}_{i}=\vec{\omega}_{1}- \vec{\alpha}_{1}-\ldots-
\vec{\alpha}_{i-1}$ are
fundamental weights of the $n$-dimensional
vector representation of $SL(n)$.

As Bowknegt McCarthy and Pilch have shown, the operators $S_{\alpha_i}$
satisfy to the q-Serre relations and realize the
representation of quantum group $U_q(n_{+})$ [24]. Therefore,
roughly speaking,
W algebra is the algebra of invariants of $U_q(n_{+})$.
$$
W \sim Inv[U_q(n_{+})] \quad .
$$

It will be our key principle in the lattice construction. Let us note
here that finding of integrals of motion in the
W-symmetric Conformal Field Theory perturbed by
relevant operator [37]
$$
\Phi=:[exp(-\alpha_{0}\phi)]:  \quad ,
$$
where $\alpha_0$ is the affine root, reduce in the bosonized
picture to the determination of  $Inv[U_q(\hat{n}_{+})]$:
$$
IM \sim Inv[U_q(\hat{n}_{+})]=Ker(S_{\alpha_0})\cap Ker(S_{\alpha_1})\cap
Ker(S_{\alpha_2}) \cap Ker(S_{\alpha_3}) \quad  ...\quad .
$$

\vglue 5pt
\line{\tenbf 1.2 Feigin's construction of lattice W algebras. ([22]). \hfil}
Let us consider an infinite set of points
$x_{-\infty}, ... , x_1, x_2, x_3, ...,x_\infty $ on the line (fig.1). One can
think of $x_i$ as a vertex operator
$$
V_\alpha=:[exp(\alpha \phi)]:
\tag 1.2.1$$
in the point $i$. Denote by $\Lambda$ the root lattice of $g$ endowed
by standard scalar product
$$
<\alpha_i,\alpha_j>=a_{ij}  \quad ,
\tag 1.2.2$$
where $(a_{ij})$ is a Cartan matrix of $g$.
Define the multigrading on $x_i$
by the map:
$$
\eqalign{
x_i &\rightarrow \Lambda \cr
deg(x_i)&=\alpha_j \quad , i \in Z, \quad  j=1,...,r.\cr
}
\tag 1.2.3$$
Imitating the exchange relations for vertex operators
$$
V_\alpha(z)  V_\beta(\zeta)=
q^{<\alpha,\beta>}V_\beta(\zeta) V_\alpha(z)         \quad ,
\tag 1.2.4$$
determine now skew polynomial algebra with basic relations:
$$
x_i x_j=q^{<deg(i),deg(j)>}x_jx_i \quad i<j .
\tag 1.2.5$$
Having in our mind to find the analogue of screening operators, put
$$
S_{\alpha_i}=\sum_{deg (x_j)=\alpha_i}x_j      \quad .
\tag 1.2.6$$
One can immediately prove the following lemma:
\vglue 5pt
\noindent
{\tenbf Lemma.} Operators $S_{\alpha_i}$ satisfy to the q-Serre relations:
$$
(ad_{S_{\alpha_i}}^{1-a_{ij}})_qS_{\alpha_j}=0 \quad .
\tag 1.2.7$$

These operators $S_{\alpha_j}$ constitute the $U_q(n_{+})$ algebra,
and formulas for comultiplication, antipod and counit are of the
form:
$$
\eqalign{
&\Delta S_{\alpha_j}= S_{\alpha_j}\otimes 1+q^{h_i}\otimes S_{\alpha_j}\quad
,\cr
&S(S_{\alpha_j})=-S_{\alpha_j} \quad ,\cr
&\epsilon(S_{\alpha_j})=0      \quad , \cr
}
\tag 1.2.8$$
while the actions of operators $h_s$ will look like
$$
\eqalign{
&h_i(P)=<degP,\alpha_i>P \quad ,\cr
&h_ih_j=h_jh_i\cr
&S(h_i)=-h_i\cr
&\Delta h_i=h_i\otimes1+1\otimes h_i \cr
&\epsilon (h_i)=0 \quad . \cr
}
\tag 1.2.9$$

Due to the eqs. (1.27)-(1.2.9) and property
$$
q^{h_i}S_{\alpha_j}=q^{a_{ij}}S_{\alpha_j}q^{h_i}
$$
operators $h_i$ and $S_{\alpha_i}$ constitute the borel part
$U_q(b_{+})$ of quantum universal enveloping algebra $U_q(g)$.

Consider now the algebra of formal Loran's series $C[x_i,x_i^{-1}]$
with
$$
deg(x_i^{-1})=-deg(x_i)
\tag 1.2.10$$
In according to the general rule the adjoint action of quantum
group $U_q(n_{+})$ is determined by q-commutation with screening
operator
$$
S_{\alpha_i}(P)=S_{\alpha_i}P-q^{<degP,\alpha_i>}PS_{\alpha_i}
\tag 1.2.11$$
We can give the following definition of lattice $W$ algebra:
\vglue 5pt
\noindent
{\tenbf Definition.} Generators of lattice $W$ algebra associated with simple
Lie algebra $g$ constitute the functional basis of space
$$
Inv_{U_q(b_{+})}(C[x_i,x_i^{-1}]) \quad .
$$
Here we added new requrements:
$$
h_i(W)=0
\tag 1.2.12$$
to have an scaling invariance $x_i\rightarrow \lambda x_i$ and to satisfy
the requirement of finiteness.
Another essential argument is that we are looking for local expression for
generators: namely, like local fields $W,T$ lattice generators must
to commute if they far enough from each other.

So the problem is to find solution of the system
of difference equations from infinite number non-commutative
variables. It is significant that commutation relations (1.2.5)
depend on the sign of the difference $(i-j)$ only. We should
try to find all solutions of the system:
$$
\eqalign{
&[S_{\alpha_i},w_1]_q=0 \quad i=1,...,r \quad ,\cr
&h_i(w_1)=<deg(w_1),\alpha_i>w_1=0 \cr
}
\tag 1.2.13$$
where $w_1=w_1(x_1,x_2,x_3,...,x_{k-1},x_k)$. Then we will obtain
the whole set of generators by shift:
$$
\eqalign{
w_2&=w_1[x_1\rightarrow x_2,x_2\rightarrow x_3,x_3\rightarrow x_4,...]\quad
,\cr
w_3&=w_1[x_1\rightarrow x_3,x_2\rightarrow x_4,x_3\rightarrow x_5,...]\quad
,\cr
&etc. \cr
}
\tag 1.2.14$$
\vglue 5pt
\line{\tenbf 2. Examples: Virasoro and $W_3$ on the lattice. \hfil}
In this section we turn our attention to the explicit construction of
lattice algebras. Let us assume for simplicity that deformation parameter $q$
is in generic position. In this case one can apply to the classical
limit $q\rightarrow 1$ and reduce rather complicated
system (1.2.13) of the difference equations
to the ordinary one. Screening operators in this limit turn to
be differential operators of first order acting on the manifold
with the coordinates $x_i$. One can easily obtain nonstandard
realization of universal enveloping algebras $U(b_{+})$ and
solve the system of ordinary differential equations. Classical
solution help us to "guess" the right answer for the quantum case.
Moreover, if we consider deformation in the generic position,
then it is possible to determine the neccesary number of variables
to find non-trivial invariant. Indeed, in the classical
case we have $dim(b_{+})$ constraints and trivial solution
$w=const$. So one can expect that nontrivial invariant
would depend on $[dim(b_{+})+1]$ variables at least.
(It is not true for special values of $q$ and non-regular maps
(1.2.3))
\vglue 5pt
\line{\tenbf 2.1 Faddeev-Takhtajan-Volkov algebra. \hfil}
Virasoro algebra is connected with $sl(2)$ algebra. In this case
we have the following basic relations:
$$
x_i x_j=q x_j x_i \quad , \quad i<j
\tag 2.1.1$$
and system of equations is:
$$
\eqalign{
&deg(\sigma)=0 \quad ,\cr
&[\sum x_i,\sigma]=0 \quad .\cr
}
\tag 2.1.2$$
As we noted before, in fact, one need to solve the following equation:
$$
(x_1+x_2+x_3)\sigma(x_1,x_2,x_3)=\sigma(x_1,x_2,x_3)(x_1+x_2+x_3)
\tag 2.1.3$$
and the solution must have zero grading. We have two obvious solutions
of this equation:
$$
\eqalign{
&x_1+x_2+x_3 \quad ,\cr
&x_1x_2^{-1}x_3 \quad ,\cr
}
\tag 2.1.4$$
and zero-grading invariant is
$$
\sigma_1=(x_1+x_2+x_3)x_1^{-1}x_2x_3^{-1}\quad .
\tag 2.1.5$$
All other basic generators of lattice Virasoro algebra
are obtained by simple shift. This algebra was found
from another point of vew by Volkov and its classical
version was appeared in the work Tahtadjan and Faddeev [25].
At the classical level lattice Virasoro has the following Poisson
brackets:
$$
\eqalign{
&\{ S_1,S_2\}=S_1S_2(1-S_1-S_2) \quad ,\cr
&\{S_1,S_3\}=S_1S_2S_3 \quad , \cr
&\{S_1,S_i\}=0, \quad |i-1|>2 \quad ,\cr
}
\tag 2.1.6$$
where
$$
S_i= {1\over \sigma_i+1}
\tag 2.1.7$$
and Poisson brackets of any $S_i$ are obtained by shift
$[1\rightarrow i], [2\rightarrow i+1], [3\rightarrow i+2] $ etc.
Faddeev and Takhtajan found this Poisson structure by studing
of Volterra system:
$$
\dot{S}_i=\{H,S_i\}=S_i(S_{i+1}-S_{i-1}) \quad ,
\tag 2.1.8$$
where hamiltonian $H$ has the form $H=\sum[\ln (S_i)]$.
\vglue 5pt
\line{\tenbf 2.2 Lattice $W_3$  algebra. \hfil}
Let us consider folowing example of lattice algebra associated to
Lie algebra $sl(3)$. There are several ways to define
the grading.
We put regular coloring of the points, exactly as in the fig.2:
$$
deg(x_{2n})=\alpha_1 \quad, \quad deg(x_{2n+1})=\alpha_2
\tag 2.2.1$$
and the commuattion relations:
$$
\eqalign{
&x_nx_{2n+k}=qx_{2n+k}x_n \quad , \quad k>0 \quad ,\cr
&x_nx_{2n+k+1}=q^{-{1\over 2}}x_{2n+k+1}x_n \quad .\cr
}
\tag 2.2.2$$
Applying to the classical limit [35] one can prove that
invariants of $U_q(sl(3))_{+}$ has the form
$$
\eqalign{
\tau_1=&(x_4x_5+x_2x_5+x_2x_3)x_6x_1(x_2x_1+x_4x_1+x_4x_2)^{-1}\cr
&(x_4x_3+x_6x_5+x_6x_3)^{-1} \quad ,\cr
\tau_2=&\tau_1(x_1\rightarrow x_2, x_2\rightarrow
x_3, x_3 \rightarrow x_4, \quad etc. ) \quad ,\cr
&etc.\cr
}
\tag 2.2.3$$
These functions from non-commutative variables determine the functional
basis in the invariant space of ${U_q(sl(3))}_{+}$.

It is rather combersome matter to find an algebra of these invariants
but
in the classical limit
$$
\{\tau_i,\tau_j\}=\lim_{q\rightarrow 1}{1 \over 1-q}[\tau_i,\tau_j] \quad ,
\tag 2.2.4$$
the calculations gives us the following result:
$$
\eqalign{
\{\tau_1,\tau_6\}&=-\tau_1\tau_3\tau_4\tau_6 \quad ,\cr
\{\tau_1,\tau_5\}&=\tau_1\tau_5[\tau_2\tau_3+\tau_3\tau_4-\tau3] \quad ,\cr
\{\tau_1,\tau_4\}&=-\tau_1\tau_4[\tau_1\tau_2+\tau_2\tau_3+\tau_3\tau_4-
\tau_2-\tau_3]\quad ,\cr
\{\tau_1,\tau_3\}&=\tau_1\tau_3[\tau_1\tau_2+\tau_2\tau_3-
\tau_1-\tau_2-\tau_3+1]\quad ,\cr
\{\tau_1,\tau_2\}&=-\tau_1\tau_2[\tau_1\tau_2-\tau_1-\tau_2+1] \quad ,\cr
\{\tau_1,\tau_i\}&=0 \quad ,\quad |i-1|>5\quad .
}
\tag 2.2.5$$
The remaining Poisson brackets are obtained through these Poisson brackets
by shifts.
Nevertheless this Poisson structure seems to be unwieldy
there is exist some diagramm which show some kind of symmetry:
$$
\{\tau_1,\tau_i\}=\tau_1\tau_i \times \Gamma_i
\tag 2.2.6$$
where $\Gamma$  is given by fig.(3-4).
(May be it is possible to write down the
Poisson brackets for $Wsl(n)$ without explicit knowledge of
lattice bosonization?)

Having such a hamiltonian structure one can define
differential-difference chain of non-linear equations:
$$
\eqalign{
H&=\sum [ln(\tau_i)]\cr
\dot{\tau_j}&=\{\tau_j,H\}=\tau_j \times \sum_i \Gamma_i\cr
}
\tag 2.2.7$$
where $\sum \Gamma_i$ denote that we should summarize all
terms in the diagramm fig.4.
For example we have:
$$
\eqalign{
\dot{\tau_1}&=\tau_1\{[-\tau_1\tau_2
+\tau_2\tau_3-\tau_3\tau_4+\tau_2+\tau_3]\cr
&-[-\tau_1\tau_{0}+\tau_{0}\tau_{-1}-\tau_{-1}\tau_{-2}+\tau_{0}+\tau_{-1}]\}\cr
&etc.\cr
}
\tag 2.2.8$$
Probably it would be interesting to investigate this
non-linear chain which seems to be integrable.
\vglue 12pt
\line{\tenbf 3. Generalization: $U_q(sl(2))$. \hfil}
In this section we turn our attention to the question what is the role
of second part of quantum group. To begin with let us consider
some local lattice field $F$ which has in the neighborhood of
point $1$ the form (fig.5):
$$
F(1)_0=F(x_1,x_2,x_3,..., x_k)=\sum C_{\{\beta\}}x_1^{\beta_1}
x_2^{\beta_2}...x_k^{\beta_k} \quad .
\tag 3.1$$
The worth important observation is that screening operator $S_{\alpha_i}$
acts on local field $F(1)_0$ only by halh of infinite sum:
$$
\eqalign{
&S_{\alpha_i}=U_{-}+\sum_1^k[x_i]+U_{+}\quad ,\cr
&U_{-}=\sum_{-\infty}^0x_i \quad ,\cr
&U_{+}=\sum_{k+1}^{\infty}x_i\quad .\cr
}
\tag 3.2$$
Namely,
$$
\eqalign{
S_{\alpha_i}(F_1)&=[x_1+x_2+x_3+...+x_k,F_1]_q+[U_{+},F_1]_q=\cr
&=[x_1+x_2+x_3+...+x_k,F_1]_q+[1-q^{2degF_1}]U_{+}F_1\quad .\cr
}
\tag 3.3$$
The expressions like $U_{+}F_1$ may be considered as the lattice
analogues of the fields dressed by screening operator (fig.6-7):
$$
\bigl(\oint{:[exp(-\alpha_{i}\phi)]:F{dz}}\bigr) \rightarrow (U_{+}F_1)
\tag 3.4$$
To investigate lattice representation of quantum group $sl_2$
if we should extend our space $C[x_i,x_i^{-1}]$ by non-local expressions
which are given by the action of screening operators:
$$
C[x_i,x_i^{-1}]\otimes U_q(sl_2)_{+} \quad .
$$
To introduce action of $U_q(sl_2)_{-}$ on this space, let us define a
q-derivation of the "screening variable" $U_{+}$:
$$
\eqalign{
&D_q \equiv {\partial \over \partial U_{+}}\quad ,\cr
&D_q(U_{+})=1 \quad ,\cr
&D_q(x_{i})=0 \quad, \quad (i \quad finite)   \quad ,\cr
&D_q(fg)=D_q(f)g+q^{(degf)}fD_q(g)  \quad . \cr
}
\tag 3.5$$
One can immediately show that operators $(D_q,h,S)$ constitute
twisted $sl(2)_q$ algebra:
$$
\eqalign{
&D_qS-qSD_q=1-q^{2h}\quad ,\cr
&q^hD_qq^{-h}=q^{-1}D_q \quad ,\cr
&q^hSq^{-h}=qS \quad .\cr
}
\tag 3.6$$
After the change
$$
D_q \rightarrow {q^{-h}D_q \over 1-q}
\tag 3.7$$
we will obtain an ordinary quantum group $sl(2)_q$.
\vglue 12pt
\line{\tenbf 4. 2D-module of $U_q(sl(2))$. Exchange algebra. \hfil}
\vglue 5pt
In previous section we received the realization of quantum group
$U_q(sl(2))$ on the lattice. One can try to investigate the
representations of quantum groups on this lattice.

If we have local feild $F_0=(..., F(1)_0, F(2)_0,...$ and modules created from
$F$ by the action of screenings then their exchange algebra
determined by universal $R$-matrix (Drineld) for $U_q(sl(2))$:
$$
\eqalign{
&F(1)_i=[S,[S,[...[S,F(1)_0] \quad (i-times)\quad ,\cr
&F(2)_j=[S,[S,[...[S,F(2)_0] \quad (j-times)\quad ,\cr
&F(1)_iF(2)_j=(R)_{ij}^{kl}F(2)_lF(1)_k \quad .\cr
}
\tag 4.1$$
Simplest possibility of such a module is two dimensional module.
Unfortunately, naive "highest weight"  $x^{-{1\over 2}}$ isn't
true becouse it creates redducible module.
One can prove, however, that proper expression for "highest weight" in
this case
is given by the expression
$$
F(1)_{-{1\over 2}}=x_1^{1\over 2}x_2^{-{1\over 2}}(x_1+x_2)^{-{1\over 2}}
\quad ,
\tag 4.2$$
while second vector in this module is given by the formula:
$$
F(1)_{1\over 2}=[S,A_{-{1\over 2}}]_q=
(1-q)U_{+}x_1^{1\over 2}x_2^{-{1\over 2}}(x_1+x_2)^{-{1\over 2}}\quad ,
\tag 4.3$$
where
$$
U_{+}=\sum_3^\infty x_i\quad .
\tag 4.4$$
By shift $x_1,x_2\rightarrow x_3,x_4$ we find another
module
$$
\eqalign{
&F(2)_{-{1\over 2}}=x_3^{1\over 2}x_4^{-{1\over 2}}(x_3+x_4)^{-{1\over 2}}\quad
,\cr
&F(2)_{1\over 2}=[S,B_{-{1\over 2}}]=(1-q)[U_{+}-x_3-x_4]
x_3^{1\over 2}x_4^{-{1\over 2}}(x_3+x_4)^{-{1\over 2}}\cr
}
\tag 4.5$$
and $R$ matrix in the representation ${1\over 2},{1\over 2}$
has the well-known form.
In this picture one can find alternative expression for
lattice Virasoro algebra. Indeed, the expressions like
$$
\Delta_{13}=F(1)_{-{1\over 2}}F(2)_{1\over 2}-q^{1\over 2}F(1)_{1\over
2}F(2)_{-{1\over 2}}
\tag 4.7$$
belong to the invariant space of $U_q(sl(2))$ and therefore have to be
invariants. Really, the condition
$$
D_q(\Delta_{13})=0
$$
denotes that this expression is local, while invariance under the action
of $S$ and $h$ is the definition of Virasoro algebra (sec.1.2).

Even without knowledge of exchange algebra one can determine the invariants
like
$$
\Delta_{12}=F(1)_{-{1\over 2}}F(1/2)_{1\over 2}-
q^{1\over 2}F(1)_{1\over 2}F({1/2})_{-{1\over 2}}
\tag 4.8$$
where (fig.8)
$$
\eqalign{
&F(1/2)_{-{1\over 2}}=x_2^{1\over 2}x_3^{-{1\over 2}}(x_2+x_3)^{-{1\over
2}}\quad ,\cr
&F(1/2)_{1\over 2}=[S,F(1/2)_{-{1\over 2}}]=(1-q)[U_{+}-x_3]
x_2^{1\over 2}x_3^{-{1\over 2}}(x_2+x_3)^{-{1\over 2}}\quad .\cr
}
\tag 4.9$$
It is interesting that 3-point invariant
$$
\Sigma=\Delta_{13}^{-1}\Delta_{12}=(x_3+x_4)^{-{1\over 2}}x_4^{1\over 2}
x_3^{1\over 2}(x_2+x_3)^{-{1\over 2}}
\tag 4.10$$
on the classical level coinsides with Tahtadjan-Faddeev
Virasoro algebra:
$$
\Sigma^{-2}=\sigma+1\quad .
\tag 4.11$$
On the quantum level I have not proved that $\Sigma$ can be expressed
through $\sigma$ and vice versa. But I think it is right.

Let us note, that 3-dimensional module is generated by
the $S$-operator action on $x^{-1}$. The exchange relation
in this case are determined through the quantum $R^{1,1}$ matrix
for $U_q(sl_2)$.

Hence we can start from exchange algebra [10] without explicit knowledge
of vectors in module and define generators of $W$ algebra as an
invariant of $U_q(g)$. The algebra of invariants gives us
$W$ lattice algebra.
\vglue 12pt
\line{\tenbf 5. Generalizations: affine case and integrals of motion.\hfil}
\vglue 5pt
\vglue 12pt
\line{\tenbf 5.1 Lattice integrals of motion [22].\hfil}
\vglue 5pt

Let now turn our attention to the integrals of motion in perturbed
conformal field theories with $W$ symmetry [37].
The integrable perturbation of such a theories
is given by the relevant field
$$
V_{\alpha_0}=:[exp(\alpha_0\phi)]:
\tag 5.1.1$$
where $\alpha_0$ is the additional (affine) root of the affinization
of Lie algebra $g$. As Zamolodchikov proved [33], the
determination of integrals of motion in perturbed conformal
field theories is reduced to the finding of the kernel of
operator
$$
S_{\alpha_{0}}=\oint{:[exp(-\alpha_{0}\phi)]:{dz}}
\tag 5.1.2$$
in the vacuum representation of $W$ algebra. Dealing with bozonizied
picture we can rewrite this problem [23] to the finding
of screening operators kernels intersection
$$
S_{\alpha_{i}}=\oint{:[exp(-\alpha_{i}\phi)]:{dz}} \quad 0<i<r.
\tag 5.1.3$$
which constitute the nilpotent part of quantum affine group $U_q(\hat{g})$.
In according to the lattice ideology, we have to
define additional screening operator
through our non-commutative variables $x_i$. Consider, for example,
$\hat{sl}(n)$ case with Cartan matrix $a_{ij}$.

Define the multigrading in this case by the regular map (fig.7):
$$
deg(x_{i[mod(n-1)]})=\alpha_i \quad  .
\tag 5.1.5$$
We have (n-1) ordinary screening operators
$$
S_i=\sum_{k=-\infty}^\infty x_{[i+k(n-1)]} \quad ,\quad i=1,...,n-1.
\tag 5.1.6$$
Our idea is to construct affine generator as following sum:
$$
S_0=\sum_{k=-\infty}^\infty
(x_{[1+k(n-1)]}x_{[2+k(n-1)]}...x_{[n-1+k(n-1)]})^{-1}
\quad
\tag 5.1.7$$
which has proper grading and corresponds in the continious limit
to the operator $V_{\alpha_0}$.
It is rather simple matter to prove the folowing lemma:
\vglue 5pt
\noindent
{\tenbf Lemma.} Operators $S_i$, ($i=0,...,n-1$) satisfy
to the q-Serre relations for the quantum affine group $U_q(\hat{sl}(n))$.
$$
(ad_{S_{\alpha_i}}^{1-a_{ij}})_qS_{\alpha_j}=0 \quad ,
\tag 5.1.8$$
where  $i\neq j; i,j=0,..,n-1 $.

Now operators $S_i$ gives us the
representation of the nilpotent part
of quantum affine Lie algebra $U_q(\hat{sl}(n)_{+})$.
Therefore, mathematically, integrals of motion problem
is defined by similar way:
\vglue 5pt
\noindent
{\tenbf Definition.} Integrals of motion on the lattice constitute
the invariant space of nilpotent part of quantum affine Lie
algebra:
$$
IM\sim Inv_{U_q(\hat{sl}(n)_{+})}(C[x_i,x_i^{-1}])
$$

Such as we have infinitely many generators of quantum affine group
$U_q(\hat{sl(n)}_{+})$
(i.e. the number of constraints is infinite) then there is no hope
to find local invariants. But we can try to find the "local density"
of "integrals". Namely, such functions, commutator of which
with screening operator is given by "total derivation":
$$
[S_i, I(x)]=D_l(P)=P(x_i,x_{i+1},x_{i+2},
...,x_{i+k})-P(x_{i+l},x_{i+1+l},x_{i+2+l},...,x_{i+k+l}) \quad , l\in Z\quad .
\tag 5.1.8$$
\vglue 12pt
\line{\tenbf 5.2 Volkov's scheme.\hfil}
\vglue 5pt
In this section we will represent Volkov's method possesing to
determine the invariant space of quantum affine group through
the solution of some system of difference equation.

For simplicity consider first example of integrals of motion
in the conformal field theory perturbed by field $\Phi_{13}$.
Such as such a field is represented by vertex operator
$$
\Phi_{13}=:[exp(-\phi)]:     \quad ,
\tag 5.2.1$$
then its resonable lattice analogue has the form $x^{-1}$.
One can immeaditely prove that
$$
(ad_{S_{\alpha_i}}^3)_qS_{\alpha_j}=0 \quad , \quad i\neq j\quad, \quad i,j=0,1
\tag 5.2.2$$
and these generators constitute $U_q(\hat{sl}(2)_{+})$ algebra.
Hence, we have two screenings in this case:
$$
\eqalign{
&S_0=\sum x_i^{-1}\quad ,\cr
&S_1=\sum x_i\quad .\cr
}
\tag 5.2.3$$
Let us consider two points $x_1$ and $x_2$. The main idea is to add
"spectral parameter" $\beta$ to the two-point screening operators and define
some analogue of "$R$" matrix:
$$
\eqalign{
&(\beta x_1+x_2)R(x_1,x_2)=R(x_1,x_2)(x_1+\beta x_2)\quad ,\cr
&(\beta x_1^{-1}+x_2^{-1})R(x_1,x_2)=
R(x_1,x_2)(x_1^{-1}+\beta x_2^{-1})\cr
}
\tag 5.2.4$$
If we could solve these equations then we construct $R_{i,i+1}$ by
simple shift of variables and the product
$$
R=\prod_{-\infty}^\infty R_{i,i+1}
\tag 5.2.5$$
(or more explicitly, logorithm of this product).
gives us the generating funtion for integrals of motion.
Let now $R_{1,2}=R_{1,2}(x_1x_2^{-1};\beta)=R_{1,2}(u;\beta)$.
Then both equations (5.2.4) are reduced
to the following linear difference equation:
$$
(\beta u+1)R_{1,2}(q^{-1}u;\beta)=(u+\beta)R_{1,2}(u;\beta)
\tag 5.2.6$$
which was appeared in the work [27]. For q in generic position one of the
the solutions
of this equation has the form:
$$
R(u,\beta)=\prod_0^\infty{1+\beta uq^{-i} \over \beta+q^{-i}u}
\tag 5.2.7$$
This expression is rather interesting: S.Kryukov notice that it
can be formally represented as the two-points correlation
function of two q-deformed bosonic fields. Moreover, in
the limit $q\rightarrow 1$ it leads to the expressions
like dilogorithms.
\vglue 12pt
\line{\tenbf 5.3 Integrals of motion. Example $\hat{sl}(3)$.\hfil}
\vglue 5pt
Let us describe direct generalization of this scheme for the case of
$\hat{sl(3)}$ algebra.

Additional screening now has the form
$$
S_{\alpha_0}=\sum_{k\in Z}[(x_{[2k]}x_{[2k+1]})^{-1}]\quad .
$$
It is easy to check by induction, that
$$
S_i^2S_j-[q^{1\over 2}+q^{-{1\over 2}}]S_iS_jS_i+S_jS_i^2=0 \quad i\neq j,
\quad i,j=0,1,2.
\tag 5.3.1$$
Correspondent difference equations system now has the form:
$$
\eqalign{
&[\beta x_1+x_3]R=R[x_1+\beta x_3]\quad ,\cr
&[\beta x_2+x_4]R=R[x_2+\beta x_4]\quad ,\cr
&[\beta {(x_1x_3)}^{-1}+{(x_2x_4)}^{-1}]R=
R[{(x_1x_3)}^{-1}+\beta {(x_2x_4)}^{-1}]\cr
}
$$
Assuming, that
$$
\eqalign{
&R=R(u_1,u_2) \quad \cr
&u_1=x_1/x_3 \quad, \quad u_2=x_2/x_4 \quad, \quad u_1u_2=q^{-1}u_2u_1\quad
.\cr
}
$$
we obtain the following system:
$$
\eqalign{
&(q\beta u_1+1)R(u_1,u_2)=qu_1R(qu_1,u_2)+\beta R(qu_1,q^{-1}u_2)\cr
&(q\beta u_2+1)R(u_1,u_2)=qu_2R(q^{-1}u_1,qu_2)+\beta R(u_1,qu_2)\cr
}
\tag 5.3.2$$

For the moment I don't know the proper way to solve this system of
difference equations from non-commutative entries, but
trivial power expansion gives us the following
solution:
$$
R_{1,2}=F(-q\beta u_1,q^{-{1\over 2}}u_2)
\tag 5.3.3$$
where $F$ is:
$$
\eqalign{
&F(x,y)=\sum{q^{m^2\over 2}({1\over \beta})_n\over
(\beta)_n(q\beta)_{n-m}}x^n y^m  \quad,\cr
&(a)_n=(1-a)(1-qa)...(1-q^{n-1}a) \quad .\cr
}
\tag 5.3.4$$
One can write down similar difference equations and solutions
for the any case $\hat{sl}(n)$
in the form of generalized q-hypergeometric Gorn's series from
non-commutative variables.

\vglue 12pt
\line{\tenbf 5.Conclusion. \hfil}
\vglue 5pt
There are many directions in this approach to be developed:

1.Consider similar picture for the lattice $W$-algebras
associated to other simple (affine) algebras.

2.Investigate similar models for more complicated
case when $q^p=1, \quad p\in Z$.

3.Consider non-regular coloring of points.

4.Construct the realization of quantum affine Lie algebras and
investigate it representations.

\vglue 12pt
\line{\tenbf Acknowledgments.\hfil}
\vglue 5pt
It is a pleasure to thank B.Feigin for suggesting
this problem to me
and enlightening discussions. I also thank
A.Belavin and members of his Seminar in Landau Institute:A.Antonov,
A.Kadeishwili, S.V.Kryukov, M.Lashkevich, S.Parchomenko,
V.Postnikov for stimulating discussions. I am grateful
to B.Enriquez for bringing to my attention work [35]

This work was supported by Landau Schoolarship awarded
by Forkschurgzentrum Julich GmbH and in part by Soros Foundation
Grant awarded by the American Physical Society.

\vglue 12pt
\line{\tenbf References \hfil}
\vglue 5pt

\medskip
\ninerm
\baselineskip=11pt
\frenchspacing

\item{1} A.B.Zamolodchikov, {\nineit Theor. Math. phys.} {\ninebf 65},
         1205 (1986).
\item{2} A.B.Zamolodchikov and V.A.Fateev, {\nineit Nucl.Phys.} {\ninebf B280
        [FS18]}, 644 (1987).
\item{3} M.Gel'fand and L.A.Dikii, preprint {\nineit IPM AN SSSR}
        {\ninebf 136}, (1978), Moscow.
\item{4} A.A.Belavin, A.M.Polyakov, A.B.Zamolodchikov, {\nineit Nucl.Phys.}
{\ninebf B241}, 333 (1984).
\item{5} S.Lukyanov, {\nineit Funct.Anal.Appl.} {\ninebf 12}, 466 (1988).
\item{6} V.A.Fateev, S.L.Lukyanov, {\nineit Intern.J.Mod.Phys.}
         {\ninebf A3}, 507 (1988).
\item{7} V.A.Fateev, S.L.Lukyanov, {\nineit Intern.J.Mod.Phys.}
         {\ninebf A7}, 853 (1992).
\item{8} V.A.Fateev, S.L.Lukyanov, {\nineit Intern.J.Mod.Phys.}
         {\ninebf A7}, 1325 (1992).
\item{9} A.Bilal and J.-L.Gervais, {\nineit Phys.Lett.}
         {\ninebf B206}, 412 (1988).
\item{10} O.Babelon, {\nineit Phys.Lett.} {\ninebf B215}, 523 (1988).
\item{11} M.Bershadsky, H.Ooguri, {\nineit Commun. Math.Phys.}
        {\ninebf 126}, 49 (1989).
\item{12} B.Feigin, E.Frenkel, {\nineit Phys.Lett.} {\ninebf B240}, 75 (1990).
\item{13} B.Feigin, E.Frenkel, Preprint {{\nineit MSRI} {\ninebf 04029-91}.
\item{14} A.A.Belavin {\nineit Adv.Stud.Pure Math.} {\ninebf 19}, 117 (1989).
\item{15} F.Bais, P.Bowknegt, K.Schoutens, M.Surridge, {\nineit Nucl.Phys.}
 {\ninebf B304}, 348 (1988).
\item{16} V.G.Drinfel'd, V.V.Sokolov, {\nineit "Mod.probl.in math."}
        Moscow VINITI, 81 (1984).
\item{17} P.Matheu, {\nineit Phys.Lett.} {\ninebf B208}, 412 (1988).
\item{18} P.Di Francesco, C.Itzykson and J.-B.Zuber,
{\nineit Sacley preprint } {\ninebf SPHT/90-149}.
\item{19} V.Yu.Ovsienko, B.A.Khesin, {\nineit Funct.Anal.Appl.}
         {\ninebf 24}, 33 (1988).
\item{20} Ya.P.Pugay, {\nineit Phys.Lett.} {\ninebf B279}, 34 (1992).
\item{21} B.Enriquez, Preprint (1992).
\item{22} B.Feigin Lectures given in Landau Inst.
\item{23} B.Feigin, E.Frenkel, Preprint {{\nineit RIMS} {\ninebf 827-91}.
\item{24} P.Bowknegt, J.McCarthy, K.Pilch, {\nineit Commun. Math.Phys.}
         {\ninebf 132}, 125 (1990).
\item{25} L.D.Faddeev, L.A.Tahtadjan, {\nineit Lect.Notes in Phys.}
        v.246, 166 (1986).
\item{26} Volkov A.Yu. {\nineit Theor.Math.Phys.} {\ninebf 74},  (1988).
\item{27} A.Yu.Volkov, L.D.Faddeev, {{\nineit Teor.Math.Phys}
          {\ninebf 92}, 207, 1992
\item{27} A.Alekseev, L.Faddeev, M.Semenov-Tian-Shansky, Preprint
          {{\nineit LOMI} {\ninebf 1991}
\item{28} O.Babelon, L.Bonora, Preprint {{\nineit SISSA/ISAS}
          {\ninebf 929/90/EP}.
\item{29} L.Bonora, Bonservizi, Preprint {{\nineit SISSA-ISAS}
          {\ninebf 110/92/EP}.
\item{30} K.Gawedsky, Preprint {{\nineit IHEP} {\ninebf 9/92/80}.
\item{31} F.Falceto, K.Gawedsky, Preprint {{\nineit IHEP} {\ninebf 1992}.
\item{32} V.G.Drinfel'd, in Proc. ICM, Berkeley, 798.
\item{33} A.B.Zamolodchikov {\nineit Adv.Stud.Pur.Math.} {\ninebf 19},  641
(1989).
\item{34} Ya.P.Pugay, {\nineit Intern.J.Mod.Phys.}, to appear.
\item{35} A.Yu.Volkov, {\nineit Phys.Lett.} {\ninebf A279}, 34 (1992).
\item{36} S.V.Kryukov, Ya.P.Pugay, preprint
         {{\nineit LANDAU} {\ninebf TMP/5/93}.
\item{37} V.A.Fateev, S.L.Lukyanov, Kiev preprint, 1988.

\vfil\supereject

\end